\documentclass[12pt]{article} \textheight 8.5in \textwidth 6.5in
\oddsidemargin 0in \topmargin -.35in \title{An Alternative to
Inflation} \author{Stefan Hollands\thanks{\tt stefan@gr.uchicago.edu}
and Robert M. Wald\thanks{\tt rmwa@midway.uchicago.edu} \\ \it Enrico
Fermi Institute and Department of Physics \\ \it University of Chicago
\\ \it 5640 S.~Ellis Avenue, Chicago, IL~60637, USA}
\begin{document}
\maketitle
\begin{abstract}

Inflationary models are generally credited with explaining the large
scale homogeneity, isotropy, and flatness of our universe as well as
accounting for the origin of structure (i.e, the deviations from exact
homogeneity) in our universe. We argue that the explanations provided
by inflation for the homogeneity, isotropy, and flatness of our
universe are not satisfactory, and that a proper explanation of these
features will require a much deeper understanding of the initial state
of our universe. On the other hand, inflationary models are
spectacularly successful in providing an explanation of the deviations
from homogeneity. We point out here that the fundamental mechanism
responsible for providing deviations from homogeneity---namely, the
evolutionary behavior of quantum modes with wavelength larger than the
Hubble radius---will operate whether or not inflation itself
occurs. However, if inflation did not occur, one must directly
confront the issue of the initial state of modes whose wavelength was
larger than the Hubble radius at the time at which they were ``born''.
Under some simple hypotheses concerning the ``birth time'' and initial
state of these modes (but without any ``fine tuning''), 
it is shown that non-inflationary fluid models
in the extremely early universe would result in the same density
perturbation spectrum and amplitude as inflationary models, although
there would be no ``slow roll'' enhancement of the scalar modes.

\end{abstract}

Issues concerning the origin of the universe and the origin of
structure in the universe are among the deepest and most fundamental
in science. In the absence of any theory of quantum gravity presently
capable of giving a local description of phenomena at or very near the
origin of the universe, it is difficult to know what questions would
be most fruitful to ask, and one cannot expect more than partial
answers to any questions that can be asked. Nevertheless, there is
much potentially to be gained by seeking to ask and answer fundamental
questions concerning the nature of the universe.

Among the most frequently posed fundamental questions are: 

\smallskip
\noindent
{\bf (1)} Why is the universe so nearly homogeneous and isotropic on
large scales, i.e., why is it so well described by a metric with
Robertson-Walker symmetry?

\smallskip
\noindent
{\bf (2)} Why is the spatial curvature of the universe so nearly zero
(and perhaps exactly zero); equivalently (assuming Einstein's
equation), why is the evolution timescale for our universe so much
greater than the fundamental timescales appearing in particle physics?

\smallskip
An image that seems to underlie the posing of these questions is that
of a blindfolded Creator throwing a dart towards a board of initial
conditions for the universe. It is then quite puzzling how the dart
managed to land on such special initial conditions of Robertson-Walker
symmetry and spatial flatness. If the ``blindfolded Creator'' view of
the origin of the universe were correct, then the only way the
symmetry (and perhaps flatness) of the universe could be explained
would be via dynamical evolution arguments. Now, dynamical evolution
arguments---in essence, the second law of
thermodynamics---successfully explain why an ordinary gas in a box
will (with overwhelmingly high probability) be found in a homogeneous
state if one examines it a sufficiently long time after the box was
filled with the gas, even though the box may have been filled by a
sloppy and careless technician who made no attempt to arrange for the
gas to be homogeneous. However, in non-inflationary models of our
universe, causality arguments alone would appear to preclude the
possibility of dynamical evolution bringing one close to
Robertson-Walker symmetry on large scales if one did not start out
with such symmetry\footnote{
It also should be noted that for a self-gravitating system such as 
our universe, dynamical evolution will normally tend to make
the system become more inhomogeneous with time (``Jeans' instability'').
Thus, the isotropy of the microwave background radiation on large scales
would truly be a puzzle if it {\it did} have time to equilibrate
on these scales.}. Thus, in non-inflationary models it does not
appear that the homogeneity, isotropy, and spatial flatness of the
universe can be explained by the same type of argument that
successfully accounts for the homogeneity of a box of gas.

Our view is that the creation of the universe is fundamentally
different from the creation of a box of gas. The ``sloppy technician''
may be a good model for the origin of a box of gas, but we see no
reason to believe in the ``blindfolded Creator'' model of the origin
of the universe. It would therefore be very surprising (and extremely
unsatisfying!) if the state of the universe were to be explained in
the same manner as the state of a gas in a box. {\em Indeed, it seems
clear that rather than seeking to use the second law of thermodynamics
or other dynamical arguments to explain how the universe arrived at
its current state starting from arbitrary initial conditions, we
should be seeking to use the (as yet to be developed) theory of
initial conditions of the universe to explain how the second law of
thermodynamics came into being} (see \cite{penrose}). Only when we
have such a theory of initial conditions will we know if there is
something left to ``explain'' related to the above two questions.

Nevertheless, inflationary theories were developed primarily to
provide a dynamical explanation of the symmetry and flatness of our
universe (as well as to provide a mechanism for diluting the presence
of magnetic monopoles that may have been created in the early
universe). If a sufficiently large, (nearly) spatially homogeneous
region has its stress energy dominated by the potential energy of a
field (which thereby effectively acts like a cosmological constant),
then that region will undergo an exponential expansion that will
enormously increase the size of this region, isotropize it \cite{w1},
and drastically reduce its spatial curvature. Inflation thereby
provides a very simple and elegant dynamical explanation of
the symmetry and flatness of the universe that overcomes the causality
obstacles to providing such an explanation.

However, despite its elegance, there are at least two significant
shortcomings to the explanation of the symmetry and flatness of the
present universe provided by inflationary theories---even if one
accepts the ``blindfolded Creator'' view of the origin of the
universe, so that a dynamical explanation of its current state is
desired/needed. First, although in inflationary models the initial
conditions needed to account for the symmetry and flatness of the
present universe certainly seem far less ``special'' than in models
without inflation, it seems clear that very ``special'' initial
conditions are nevertheless needed in order to enter an era of
inflation. To see this in a graphic manner, it is useful to consider a
universe that eventually collapses to a final, ``big crunch''
singularity. As the ``big crunch'' is approached, it seems
overwhelmingly improbable---and, indeed, in apparent blatant
contradiction with the second law of thermodynamics---that the matter
in the universe would suddenly coherently convert itself to scalar
field kinetic energy in just such a way that the scalar field would
``run up a potential hill'' and remain nearly perfectly balanced at
the top of the hill for a long period of exponential contraction of
the universe. In other words, it seems overwhelmingly improbable that
a collapsing universe would undergo an era of ``deflation'' just
before the ``big crunch''. Thus, the region of ``final data space''
that corresponds to a universe that did not deflate should have much
larger measure than the region corresponding to a universe that did
deflate. But the time reverse of a collapsing universe that fails to
deflate is, of course, an expanding universe that fails to
inflate. Thus, this argument strongly suggests that the region of
initial data space that fails to give rise to an era of inflation has
far larger measure than the region that does give rise to an
inflationary era, i.e., it is overwhelmingly unlikely that inflation
will occur\footnote{A more precise version of this argument can be given as 
follows: Let $\mathcal U$ be the collection of universes that start 
from a ``big bang'' type of singularity, expand to a large size and recollapse
to a ``big crunch'' type of singularity. Let $\mathcal I$ denote the space
of initial data for such universes and let $\mu_{\mathcal I}$ denote
the measure on this space used by the ``blindfolded Creator''. Let 
$\mathcal F$ denote the space of final data of the universes in $\mathcal U$, 
and let $\mu_{\mathcal F}$ denote the measure on $\mathcal F$ obtained 
from $\mu_{\mathcal I}$ via the ``time reversal'' map. Suppose that 
$\mu_{\mathcal I}$ is such that dynamical evolution from $\mathcal I$ to 
$\mathcal F$ is measure preserving (``Liouville's theorem''). Then the 
probability that a universe in $\mathcal U$ gets large by undergoing 
an era of inflation is equal to the probability that a universe in $\mathcal
U$ will undergo an era of ``deflation'' when it recollapses.    
}. We do not know the measure on the dartboard used
by the blindfolded Creator, so it does not seem possible to make this argument
quantitatively precise. But, suppose that the probability of the dart
landing on initial conditions directly giving rise to a nearly flat
Robertson-Walker model without inflation were, say, $10^{-10^{10}}$ or
smaller, whereas the probability of the dart landing on initial
conditions leading to inflation---and, thereby, to a nearly flat
Robertson-Walker universe---were, say, $10^{-10}$. Then inflation
would indeed successfully enhance the probability of creation of a
universe that looks like our universe by a factor of $10^{10^{10}}$ or
more. But could it really be said that inflation has accounted for the
creation of a universe that looks like ours?

A possible way to counter the above argument is to note that one does
not need the entire universe to undergo an era of inflation, but only
a sufficiently large portion of it. Although the probability that a
given region will inflate may be small, if this probability is
non-zero and the universe is infinite (or if infinitely many universes
are created), then {\it some} regions will inflate\footnote{
However, the argument of the preceeding footnote still applies, i.e., 
the probability of inflation should still be the same as the probability
of ``deflation''.}. Anthropic
arguments can then be invoked to explain why we happen to live in a
portion of the universe that had undergone an era of inflation.
However, we feel that it is legitimate to ask whether arguments of this
nature should be considered as belonging to the realm of science. Such
arguments are based on an assumed knowledge of quantities---such as
the probability measure on the blindfolded Creator's dartboard and the
probability of producing intelligent life in universes very different
from ours---that we have no hope of accessing at the present time and
that may well turn out to be meaningless when we have attained a
deeper understanding of nature. It is far from clear what is really
being ``explained'' by such arguments and whether, even in principle,
any nontrivial testable predictions can be made.

A second difficulty arises when one considers the details of the
models that give rise to inflation. We do not find it unreasonable to
postulate that in the very early universe there was
an era when the energy density of the universe was dominated
by the self-interaction potential energy, $V(\phi)$, of a scalar field
$\phi$. However, in order to have a sufficiently long era of inflation
from which the universe can exit in an acceptable manner, $V(\phi)$ must be
extremely flat. Additional significant constraints arise from the
quantum fluctuations in energy density produced in the inflationary models
(see below). Consequently, although scalar field models do exist that
result in inflation---at least, with suitable initial conditions, as discussed
above---one must ``tune'' the parameters in these models quite
carefully to satisfy all of the constraints \cite{st}. Thus, although
inflationary models may alleviate the ``fine tuning'' in the choice of
initial conditions, the models themselves create new ``fine tuning''
issues with regard to the properties of the scalar field\footnote{
These fine tuning difficulties with the models are alleviated in the 
chaotic inflation scenario.}.

Thus, even if one were to accept the blindfolded Creator view of the
origin of the universe, it is our view that inflationary models are not
very successful with regard to providing answers to questions (1) and
(2) above. However, the situation changes dramatically when one
considers another fundamental question:

\smallskip

\noindent
{\bf (3)} Given that the universe has nearly Robertson-Walker
symmetry, how did the departures from this symmetry originate?

\smallskip

Here inflationary models provide a very simple, natural, and beautiful
answer to this question: The departures from homogeneity arose from
the quantum fluctuations of the field responsible for inflation.
Although the overall amplitude of the density fluctuations produced by
inflation depends upon the details of the particular model (and thus
plays more the role of a constraint on inflationary models rather than
a prediction of inflation), inflationary models natually yield a
so-called ``scale free'' spectrum of density perturbations (see
below). This prediction of a scale-free spectrum has been
spectacularly confirmed during the past year by high precision
measurements of the cosmic microwave background \cite{cmb}.

The basic mechanism by which inflationary models give rise to
macroscopically important fluctuations at long wavelengths can be seen
by considering the simple model of a free, massless, minimally coupled
scalar field, $\phi$, in a spatially flat background Robertson-Walker
spacetime,
\begin{equation}
ds^{2} = - dt^2 + a^2(t) [dx^2 +dy^2 +dz^2] .
\label{RW}
\end{equation}
If we consider a plane wave mode of coordinate wavevector $\vec{k}$,
\begin{equation}
\phi(t,\vec{x}) = \phi_k(t) e^{i\vec{k} \cdot \vec{x}}
\label{k}
\end{equation}
then $\phi_k$ satisfies
\begin{equation}
\frac{d^2\phi_k}{dt^2} + 3H\frac{d\phi_k}{dt} + \frac{k^2}{a^2} \phi_k = 0,
\label{phik}
\end{equation}
where $H = a^{-1} da/dt$ is the Hubble constant. This is identical in
form to the harmonic oscillator equation with a unit mass, a
(variable) spring constant $k^2/a^2$, and a (variable) friction
damping coeficient $3H$. Consequently, when the (proper) wavelength,
$a/k$, of the mode is much smaller than the Hubble radius, $R_H =
1/H$, the mode will behave like an ordinary harmonic oscillator, with
negligible damping. On the other hand, when the wavelength is much
larger than the Hubble radius, the mode will behave like an overdamped
oscillator; its ``velocity'', $d\phi_k/dt$, will rapidly decay towards
zero and its amplitude will effectively ``freeze''.

In the quantum theory of the scalar field $\phi$, each mode $\phi_k =
(2\pi)^{-3/2} \int \exp(- i\vec{k} \cdot \vec{x}) \phi d^3x$ acts as
an independent harmonic oscillator, with Lagrangian
\begin{equation}
L_k = \frac{a^3}{2}[|d\phi_k/dt|^2 - \frac{k^2}{a^2} |\phi_k|^2],
\label{L}
\end{equation}
where the factor of $a^3$ arises from proper volume element in the
Klein-Gordon Lagrangian for $\phi$. (Note that $\phi_k$ was defined
using the coordinate volume element rather than the proper volume
element in order to obtain this simple form for $L_k$.) At a fixed
time $t$, the ground state of the oscillator defined by eq.~(\ref{L})
is a Gaussian wavefunction in $\phi_k$, with spread given by
\begin{equation}
(\Delta \phi_k)^2 = \frac{1}{2a^3(k/a)}
\label{dp}
\end{equation}
(see, e.g., eq.~(2.3.34) of \cite{s}). Now, if the proper wavelength of
the mode is much smaller than the Hubble radius, the ground state will
evolve adiabatically, and eq.~(\ref{dp}) will continue to hold at later
times. At the other extreme, if the proper wavelength of the mode is
much larger than the Hubble radius, the oscillator will be overdamped,
and the fluctuation amplitude $\Delta \phi_k$ will remain constant
with time.

It should be noted that during a ``normal'' era of evolution of the
universe (when $P \geq 0$---or, more generally, $P > - \rho/3$ where
$P$ is the pressure and $\rho$ is the mass density), the Hubble radius
will grow more rapidly than $a$, so the Hubble radius will tend to
``overtake'' the proper wavelength of modes. Thus, $\phi_k$ may evolve
from an overdamped oscillator to an underdamped oscillator, but not
vice-versa. On the other hand, during an era of inflation (when $P = -
\rho$), the Hubble constant is truly constant, whereas $a$ grows
exponentially with $t$. Thus, the proper wavelength of modes will tend
to rapidly overtake the Hubble radius.

The basic mechanism by which inflation produces a spectrum of density
perturbations appropriate to account for the origin of structure in
our universe may now be explained. In inflationary models, the modes
relevant to cosmological perturbations are assumed to be ``born'' in
their ground state at a time when their proper wavelength is much less
than the Hubble radius. These modes initially evolve adiabatically
(remaining in their ground state), so the precise time at which they
came into existence is not important. However, during an era of
inflation, their proper wavelength becomes much larger than the Hubble
radius, and their fluctuation amplitude essentially freezes at the
value
\begin{equation}
(\Delta \phi_k)^2 \sim \frac{1}{2a_0^3(k/a_0)},
\label{dp2}
\end{equation}
where $a_0$ is the value of the scale factor at the time the mode
``crossed'' the Hubble radius, i.e., at the time when
\begin{equation}
k/a_0 = H_0
\label{cross}
\end{equation}
where $H_0$ is the Hubble constant during the inflationary era. Now
consider these modes at a later time---but early enough that all of
the cosmologically relevant modes still have wavelength larger than
the Hubble radius.  Combining eqs. (\ref{dp2}) and (\ref{cross}), we
see that the fluctuation spectrum for these modes is given by
\begin{equation}
(\Delta \phi_k)^2 \sim \frac{H_0^2}{k^3},
\label{dp3}
\end{equation}
which corresponds to a ``scale free'' spectrum\footnote{\label{sf} The
normalization of the power spectrum commonly used elsewhere differs
from our conventions by a factor of $k^{-3}$ as a consequence of the
use of the volume element $dk/k$ rather than $k^2 dk$ in the inverse
Fourier transform. Thus, eq.~(\ref{dp3}) corresponds to a power
spectrum that is independent of $k$ in the alternate conventions.}. Note
that eq.~(\ref{dp3}) differs from eq.~(\ref{dp}) by a factor of
$(a/a_0)^2$, which is enormous for the modes of interest and thereby
accounts for how quantum fluctuations can have macroscopically
relevant cosmological effects.

In order for the above initial fluctuation spectrum of $\phi_k$ to
produce a corresponding initial fluctuation spectrum of the density
perturbations, it is necessary that the scalar field also make a
large, essentially classical contribution to the stress-energy of the
universe. If it does so, then the cross-terms in the stress-energy
tensor of the scalar field between the classical, homogeneous
background field $\phi_0$ and the quantum fluctuations of the scalar
field will give rise to cosmologically relevant density
perturbations. In standard inflationary models, the initially large,
background, classical energy of the scalar field is provided by
potential energy, with an extremely ``flat'' potential. The
stress-energy associated with this potential provides an effective
cosmological constant in Einstein's equation, which self-consistently
drives the evolution of the universe into an inflationary era. A
sufficiently ``slow roll'' down this potential provides a sufficiently
long era of inflation for the relevant modes to behave as described
above. An essentially\footnote{The inflationary models actually
predict logarithmic corrections to the scale-free spectrum that depend
upon the details of the model~\cite{st}, \cite{mfb}, due to the fact
that the Hubble constant is not strictly constant as the scalar field
rolls down the potential hill.}  scale-free spectrum of density
perturbations is thereby produced. The complete analysis of the
scalar density perturbations produced during inflation \cite{bst}, \cite{mfb}
is more complicated than the above analysis of a test scalar field in a
fixed, background spacetime, since, in particular, one must consider
the evolution for perturbations of the full, coupled
Einstein-scalar-field system and follow these perturbations through
reheating. Nevertheless, the basic explanation of the origin of the
scalar perturbations and their scale-free nature is as given above for a test
scalar field. 

The main purpose of this paper is to point out that the above basic mechanism
responsible for producing density perturbations in inflationary
models---namely, the evolutionary behavior of quantum modes with
wavelength larger than the Hubble radius---will operate whether or not
inflation actually occurs. Therefore, it may not be necessary to
assume that an era of inflation actually occurred in order to account
for the origin of structure in much the same way as in inflationary
models.

In the analysis of the density perturbation spectrum arising in
non-inflationary models, the modes of cosmological interest have
proper wavelength much larger than the Hubble radius throughout the
evolutionary history of the early universe. Therefore, one must
directly confront the issue of the initial state of these modes. This
issue, of course, also arises in inflationary models \cite{bm}, but
since in inflationary models the modes at early times had proper
wavelength smaller than the Hubble radius, it seems natural to
assume---as we did above---that the modes are ``born'' in their ground
state. As mentioned above, the results are then not sensitive to the
precise time at which it is assumed that the modes are born. Thus, the
issue of the initial state of modes has not played a central role in
analyses of inflationary models, since one merely needs to assume that
the modes were born in their ground state at some point prior to or
during inflation.

However, in non-inflationary models, the predictions for density
fluctuations will depend sensitively on assumptions about the initial
conditions of the modes. The assumptions that would appear most
natural concerning the time at which modes are ``born'' and their
initial state at birth depend primarily upon one's view of the validity of a
semiclassical description of our universe. It is usually assumed that
a semiclassical description will break down---and a complete theory of
quantum gravity will have to be used---if one tries to describe
phenomena on a spatial scale smaller than the Planck length, $l_P$. It
is similarly assumed that a semiclassical description will break down
in the description of phenomena occurring on a timescale smaller than
the Planck time, $t_P$. Otherwise, it is normally assumed that a
semiclassical description will, in general, be valid. However, it
should be noted that the arguments for these views do not go much
beyond the dimensional analysis given by Planck over a century
ago. Furthermore, one cannot give a Lorentz invariant version of these
criteria unless one also takes the view that the semiclassical
description breaks down for all phenomena involving null related
events, even if they are ``macroscopically separated''.

In the above conventional view, a valid semiclassical description of
spacetime structure, matter fields, and their quantum fluctuations at
all spatial scales larger than $l_P$ should suddenly become possible
at the Planck time, $t_P$. In this view, it would seem natural to
assume that all the modes with wavelengths greater than $l_P$ would be
instantaneously ``born'' at time $t_P$. As the universe expands and
these modes attain larger proper wavelengths, new modes would then
have to be continuously created at the Planck scale to ``fill in'' the
``gap'' produced by the expansion of the original modes. Now, as
already indicated above, in a non-inflationary model, the modes of
cosmological interest have a proper wavelength much larger than $l_P$
at the Planck time. If it is assumed that these modes are born in
their ground state at the Planck time, then the evolutionary era
during which their proper wavelength remains larger than the Hubble
radius will be too short to produce large enough effects to be of
cosmological interest. In addition, since the relevant
modes of different wavelengths are all created simultaneously, the
power spectrum will remain that of the ground state (i.e., $(\Delta
\phi_k)^2 \propto 1/k$) rather than the scale-free spectrum obtained
in inflationary models.

We wish to propose here an alternative view on the creation of modes: We
propose that semiclassical physics applies (in some rough sense) to
phenomena on spatial scales larger than some fundamental length,
$l_0$, which, presumably, is of order the Planck scale or, perhaps,
the grand unification scale. In this view, it would make sense to talk
about a classical metric and quantum fields at times nominally earlier
than the Planck time\footnote{
In other words, we suggest that it makes sense to talk about 
phenomena ``emerging out of the spacetime foam'' at length scales
greater than $l_0$, and that {\it some} sort of semiclassical description
of such phenomena may be possible even in an era that would correspond
to $t \ll t_P$ in the naive extrapolation of a semiclassical solution 
of Einstein's equations to early times. Note however that we do {\it not} suggest that 
an accurate semiclassical description should be given by such a 
naive extrapolation of the classical spacetime metric to time 
earlier than $t_P$. Instead, the ``correct'' semiclassical description that we 
have in mind would presumably be obtained by some suitable ``coarse graining''
of the degrees of freedom of quantum gravity over length scales $< l_0$. 
}, provided that one restricts consideration to
phenomena occurring on spatial scales larger than $l_0$. Thus, in this
view, it would be natural to treat the modes as effectively being
``born'' at a time when their proper wavelength is equal to the
fundamental scale\footnote{
A model in which modes are ``created'' when the wavelength is at 
a given spatial scale is considered in~\cite{bh}; see also~\cite{hs}.
}, $l_0$. Consequently, in this view, all of the modes
would, in effect, be continuously created over all time, in contrast
with the view that most of the modes (including all of the
cosmologically relevant ones) are created at the Planck time and the
rest are continuously created at later times.

If we assume that the modes are created in their ground state, then it
is easy to see that the calculation of the fluctuation spectrum for a
free, massless scalar field becomes identical to the inflationary
calculation sketched above, with the Hubble radius, $1/H_0$, at the time
of inflation replaced by $l_0$. Thus, the desired scale free
spectrum of an appropriate amplitude will be obtained. It should be
emphasized that to obtain this result for a scalar field in a fixed
background Robertson-Walker spacetime, no assumptions need to be made
concerning the detailed behavior of the scale factor $a(t)$ in the
early universe.

In order to construct a non-inflationary model in which density
perturbations of the desired spectrum and amplitude are produced, it
is necessary to consider a situation where there is a large background
stress-energy that is linearly perturbed by quantum fluctuations of
the appropriate spectrum and amplitude. As a simple model, suppose
that the matter in the early universe can be described on spatial
scales greater than $l_0$ by a fluid with equation of state $P = w
\rho$, where $w$ is a constant with value in the range $0 < w \leq
1$. (The case of greatest physical relevance would presumably be $w =
1/3$, but we prefer to admit a general $w$ in order to emphasize that
our results to not depend sensitively on the details of the equation
of state.) We {\em assume} that the universe is well described by a
flat Robertson-Walker model---as we await a theory of initial
conditions that might provide some kind of deeper understanding of why
this is so. To analyze the density perturbation spectrum that would be
arise in this model, we must quantize the perturbations of the coupled
Einstein-fluid system. The required analysis of this system has been
given in~\cite{mfb}, where a Lagrangian was obtained for a gauge
invariant ``velocity potential'' $v$, defined by eq.~(10.43a) of that
reference. If we define $\psi = v/a$ and transform from the
``conformal time'' variable used in~\cite{mfb} to proper time, the
action given by eq.~(10.62) of~\cite{mfb} corresponds in the case of a
$P = w \rho$ fluid to the Lagrangian
\begin{equation}
L_k = \frac{a^3}{2}[|d\psi_k/dt|^2 - \frac{c^2_s k^2}{a^2} |\psi_k|^2],
\label{L2}
\end{equation}
where $c_s = w^{1/2}$ denotes the speed of sound in the fluid. This is
precisely the same Lagrangian as for a test scalar field discussed
above, except that the sound speed, $c_s$, has replaced the speed of
light, $c=1$. We may therefore immediately write down the power
spectrum for $\psi$, valid at all later times at which the modes still have
wavelength larger than the Hubble radius
\begin{equation}
(\Delta \psi_k)^2 = \frac{1}{2a_0^2 c_s k},
\label{dp6}
\end{equation}
where $a_0$ denotes the scale factor at which the mode was
created. Under the hypothesis stated above, we take $a_0$ to be given
by $a_0 \sim k l_0$. This yields the power spectrum
\begin{equation}
(\Delta \psi_k)^2 \sim \frac{1}{2l_0^2 c_s k^3}.
\label{dp4}
\end{equation}

The gauge invariant gravitational potential $\Phi$ of \cite{mfb}
(equal to the potential $-\Phi_H$ of \cite{b} and directly related to
the ``gauge invariant fractional density perturbation'' $\delta
\epsilon/\epsilon_0$ of~\cite{mfb}) is given in terms of $\psi$ by
\begin{equation}
\Phi_k = \frac{\sqrt{3(6w+5)}}{2\sqrt{2}} \frac{l_P}{c_s}
\frac{a^2}{k^2} H \frac{d \psi_k}{dt}
\label{Phi}
\end{equation}
(see eq.~(12.8) of~\cite{mfb}; following their conventions, we have defined
$l_P^2 = \frac{8\pi}{3} \frac{G \hbar}{c^3}$). 
In order to evaluate $d\psi_k/dt$, we
integrate the equation of motion
\begin{equation}
\frac{d}{dt} \left(a^3 \frac{d\psi_k}{dt}\right) = - c^2_s k^2 a \psi_k
\label{eom}
\end{equation}
and use the fact that $\psi_k$ itself is approximately constant.
Substituting the background solution for a $P = w \rho$ fluid (namely
$a \propto t^{2/(3w +3)}$), we obtain
\begin{equation}
\frac{d\psi_k}{dt} \sim \frac{a^3_0}{a^3} \left(\frac{d\psi_k}{dt}
\right)_{t=t_0} - c^2_s
t \frac{3w+3}{3w+5} \frac{k^2}{a^2} \psi_k,
\label{eom2}
\end{equation}
where $t_0$ denotes the ``creation time'' of the mode. The first term
on the right side is negligible for $t \gg t_0$ with our assumed ground
state initial conditions for $\psi_k$. Substitution into
eq.~(\ref{Phi}) and again using the background solution yields
\begin{equation}
\Phi_k \sim - \sqrt{\frac{3}{2}} \frac{\sqrt{6w+5}}{3w+5} l_P c_s \psi_k.
\label{Phi2}
\end{equation}
Consequently, eqs.~(\ref{dp4}) and~(\ref{Phi2}) imply that the power spectrum
for modes with wavelength greater than the Hubble radius is given by
\begin{equation}
(\Delta \Phi_k)^2 \sim \frac{l^2_P}{l_0^2} \frac{3w^{1/2}(6w+5)}{4(3w+5)^2} 
\frac{1}{k^3}.
\label{dp5}
\end{equation}
This corresponds to the desired scale free spectrum of density
perturbations.  Furthermore, the correct amplitude is
obtained\footnote{Similar results to those we have just derived can be
obtained from eqs.~(12.29) and~(12.32) of \cite{mfb}, with the
understanding that the initial time appearing in eq.~(12.29) for a
given mode is now to be taken to be the time at which $k/a_0 = 1/l_0$,
rather than some fixed time that is independent of $k$. However, it
should be noted that the initial conditions chosen in \cite{mfb}
correspond roughly to a ground state condition for $v_k$ with respect
to conformal time, whereas our initial conditions correspond to a
ground state condition for $\psi_k$ with respect to proper time. This
makes a very important difference for the case $w = 1/3$, where the
initial conditions of \cite{mfb} are chosen so as to yield no effect
(i.e., a power spectrum at late times equal to a ground state
spectrum) whereas our initial conditions yield as large an effect for
$w = 1/3$ as for other values of $w$.} if we choose $l_0$ to be of
order the grand unification scale (i.e., $l_p/l_0 \sim 10^{-5}$).

Note that the amplitude of the density perturbation spectrum for 
``scalar modes'' in inflationary models differs from our formula 
eq.~(\ref{dp5}) with $l_0 = 1/H_0$ primarily in that there is no 
``slow roll enhancement factor'' in our formula. Therefore, in our
model, the amplitude of scalar modes would be suppressed 
relative to inflationary models with $H_0$ taken to be $1/l_0$.
On the other hand, the amplitude of ``tensor modes'' in our 
model should be essentially the same as that of inflationary models
with $H_0 = 1/l_0$. Consequently, our model should predict a 
larger ratio of tensor to scalar perturbations than typical inflationary
models. 

The above fluid model of the extremely early universe is undoubtedly
far too simplistic to be taken seriously as a realistic description of
phenomena occurring during that era. The above hypotheses concerning
the validity of a semiclassical description and the birth of modes are
also undoubtedly too simplistic---although not necessarily more
simplistic than conventional assumptions that would, in effect,
postulate that all of the relevant modes are born in their ground
state at the Planck time. Thus, the above model is not intended to
represent an accurate account of the origin of density fluctuations
but rather is being proposed in the spirit of an ``existence proof''
for robust alternatives to inflation.

In summary, we have argued that inflation does not satisfactorily
``solve'' the homogeneity/isotropy and flatness ``problems''---nor is
any other dynamical mechanism likely to give a satisfactory
explanation of the homogeneity/isotropy and spatial flatness of our
universe. Rather, a much deeper understanding of the nature of the
birth of our universe undoubtedly will be required. On the other hand,
the mechanism for producing density fluctuations in inflationary
models via the dynamical behavior of quantum modes with wavelength
larger than the Hubble radius provides a very simple and natural
explanation for the observed departures from homogeneity and isotropy
in our universe. This mechanism will operate whether or not inflation
occurred, but in non-inflationary models, the results will depend
crucially on one's assumptions concerning the birth of modes.
Consequently, in non-inflationary models, we are placed in a much more
uncomfortable position with regard to making reliable
predictions---although this does not mean that Nature would share our
discomfort to the degree that She would thereby choose an inflationary
model over a non-inflationary one! We have shown above that under
suitable assumptions concerning the birth of modes, a density
fluctuation spectrum for non-inflationary models can obtained (without
any ``fine tuning'') that is of the same nature as that of
inflationary models. Thus, we have provided an alternative to
inflation. However, the determination of whether this alternative is
correct (or even viable) will require a much deeper understanding than
we presently possess of the nature of the universe at and near its
birth.

We wish to thank Bill Unruh for helping to explain to us the manner in
which inflation produces density fluctuations, and we wish to thank
Sean Carroll for reading the manuscript and providing us with a number
of useful comments. This research was supported in part by NSF grant
PHY00-90138 to the University of Chicago.


\begin{thebibliography}{99}
\bibitem{penrose} R. Penrose, in \emph{General Relativity, an Einstein
Centennary Survey}, ed. by S.W. Hawking and W. Israel, Cambridge
University Press (Cambridge, 1979).
\bibitem{w1} R.M. Wald, Phys. Rev. \textbf{D28},
2118 (1983).
\bibitem{st} P.J. Steinhardt and M.S. Turner, Phys. Rev.
\textbf{D29}, 2162 (1984).
\bibitem{cmb}
C.~B.\ Netterfield {\it et al.}, astro-ph/0104460;
C. Pryke {\it et al.}, astro-ph/0104490; 
R. Stompor {\it et al.}, astro-ph/0105062 (2001).
\bibitem{s} J.J. Sakurai, \emph{Modern Quantum Mechanics}, Addison
Wesley (Reading, MA, 1994).
\bibitem{bst} J.M. Bardeen, P.J. Steinhardt and M.S. Turner,
Phys. Rev.  \textbf{D28}, 679 (1983).
\bibitem{mfb} V.F. Mukhanov, H.A. Feldman, and R.H. Brandenberger,
Phys. Rep.  \textbf{215}, 203 (1992).
\bibitem{bm} J. Martin and R.H. Brandenberger, astro-ph/0012031;
A. Kempf and J.C. Niemeyer, Phys. Rev. \textbf{D64}, 103501 (2001).
\bibitem{bh} R.~Brandenberger and P.~M.~Ho, hep-th/0203119.
\bibitem{hs} S.~F.~Hassan and M.~S.~Sloth, hep-th/0204110.
\bibitem{b} J.M. Bardeen Phys. Rev.  \textbf{D22}, 1882 (1980).
\end{thebibliography}
\end{document}